\documentclass[review]{elsarticle}

\usepackage{amsmath,amssymb,amsfonts}
\usepackage{algorithmic}
\usepackage{graphicx}
\usepackage{textcomp}
\usepackage{xcolor}
\usepackage[ruled,linesnumbered]{algorithm2e} 
\usepackage[nolist]{acronym}
\usepackage{soul}
\usepackage{marginnote}
\usepackage{url}
\usepackage[inline]{enumitem}
\usepackage{tabularx}
\usepackage{amsthm}
\usepackage{tikz}
\usepackage[utf8]{inputenc}
\usepackage[T1]{fontenc}

\newcommand{\myssec}[2]{%
  \subsection{#1}%
  \label{sec:#2}%
}
\newcommand{\rsec}[1]{%
  Sec.~\ref{sec:#1}%
}

\newcommand{\added}[1]{%
  {\color{blue}#1%
  }%
}
\renewcommand{\added}[1]{#1}

\newcommand{\myfigeps}[3][width=\columnwidth]{%
\begin{figure}[htbp!]%
\centering%
\includegraphics[#1]{figures/#2}%
\caption{#3}%
\label{fig:#2}%
\end{figure}%
}
\newcommand{\myfigdoubleeps}[5][width=\columnwidth]{%
\begin{figure*}[tb]%
\centering%
\includegraphics[#1]{figures/#2}\\
\includegraphics[#1]{figures/#3}%
\caption{#5}%
\label{fig:#4}%
\end{figure*}%
}

\newcommand{\rfig}[1]{Fig.~\ref{fig:#1}}

\newcommand{\req}[1]{Eq.~(\ref{eq:#1})}

\newenvironment{myinlinelist}%
{%
\begin{enumerate*}[label=(\roman*)]%
}%
{%
\end{enumerate*}%
}

\newenvironment{myitemlist}%
{%
\begin{itemize}[parsep=0em,leftmargin=*,label={--}]%
}%
{%
\end{itemize}%
}

\newenvironment{myenumlist}%
{%
\begin{enumerate}[parsep=0em,leftmargin=*,label=\arabic*.]%
}%
{%
\end{enumerate}%
}

\begin{document}

\title{%
  Service Differentiation and Fair Sharing\\in Distributed Quantum Computing
}

\author[1]{Claudio Cicconetti\corref{cor1}}%
\ead{c.cicconetti@iit.cnr.it}

\author[1]{Marco Conti}
\ead{m.conti@iit.cnr.it}

\author[1]{Andrea Passarella}
\ead{a.passarella@iit.cnr.it}

\cortext[cor1]{Corresponding author}
\address[1]{IIT, National Research Council, Pisa, Italy}

\begin{abstract}
  In the future, quantum computers will become widespread and a network of quantum repeaters will provide them with end-to-end entanglement of remote quantum bits.
As a result, a pervasive quantum computation infrastructure will emerge, which will unlock several novel applications, including distributed quantum computing, that is the pooling of resources on multiple computation nodes to address problem instances that are unattainable by any individual quantum computer.
In this paper, we first investigate the issue of service differentiation in this new environment.
Then, we define the problem of how to select which computation nodes should participate in each pool, so as to achieve a fair share of the quantum network resources available.
The analysis is performed via an open source simulator and the results are fully and readily available.
\end{abstract}


\maketitle

  \section{Introduction}%
  \label{sec:introduction}%
  \ac{QC} exploits the
properties of matter at very small scale to solve some problems
much faster than a classical counterpart.
Even though \ac{QC} has been theorized 40 years
ago~\cite{preskill_quantum_2021}, only recently the technology
evolution and a spur of investments have made it possible to obtain
practical results and speculate about approaching mass
deployments~\cite{sevilla_forecasting_2020}.
\ac{QC} is being already used in the chemical and pharmaceutical
industry, while new applications are being progressively unlocked in
material science, \ac{ML} and engineering optimization, production
and logistics, post-quantum security
\cite{quantum_technology_and_application_consortium__qutac_industry_2021}.
Essentially, the computational advantage of \ac{QC} stems from the
properties of superposition and entanglement of the \textit{qubits}
(i.e., the ``quantum bits''):
\begin{myinlinelist}
\item \textit{superposition}, which means that a qubit can be in a combination
of multiple states at the same time; and

\item \textit{entanglement}, which is a property exhibited by a set of
qubits that maintain their correlation even separated in space or time.
\end{myinlinelist}
%

We can expect that the computational power of a single \ac{QC} will
remain relatively limited in the near future, due to scalability
issues in maintaining a very stable and controlled environment to
cope with the flimsy nature of qubits.
On the other hand, the realization of the Quantum Internet is progressing steadily~\cite{gyongyosi_advances_2022}, with the long-term goal to enable the entanglement of qubits that reside in \acp{QC} across geographical distances.
With the diffusion of \acp{QC} and their gradual interconnection via quantum networks, a pervasive infrastructure will therefore materialize, with the potential to combine opportunistically resources from multiple \acp{QC} for the execution of specialized algorithms in a distributed fashion.
A general framework for such \textit{distributed quantum computing} has been proposed in~\cite{parekh_quantum_2021}, where the authors propose practical examples, e.g., a quantum version of the k-means clustering, which is used in unsupervised \ac{ML}.
%

A preliminary analysis of the allocation of resources among multiple quantum computers based on the characteristics of the underlying quantum network has been presented in~\cite{cicconetti_resource_2022}.
In the same work, we have also proposed a practical solution inspired by a well-known algorithm in classical data networks, i.e., \ac{DRR}~\cite{shreedhar_efficient_1995}, which we have evaluated through
simulations.
We have found that some fundamental properties of quantum networks immensely impact on the provisioning of resources, which calls for new research in this area.
This is especially manifest when considering networks of first-generation (1G) quantum repeaters~\cite{muralidharan_optimal_2016}, which do not have error-correction capabilities and are expected to be next in line for the industrialization and mass deployment in the following years~\cite{wang_field-deployable_2022}.

The contribution of this paper is twofold.
\begin{myenumlist}
    \item We evaluate the performance of the resource allocation algorithm proposed in~\cite{cicconetti_resource_2022} with differentiated services coexisting within the same quantum network.
    Furthermore, we do so by comparing the performance with two alternative algorithms, inspired by equivalents in classical problems with similar settings.
    This extends and completes the preliminary analysis in our previous work.
    
    \item We define a new problem related to fair share of resources in a quantum network among multiple applications wishing to perform distributed \ac{QC}: how to best choose the peers among those available?
    After introducing a mathematical formulation of the problem, we propose a greedy approximation algorithm, which is then evaluated thoroughly and compared to two alternatives.
\end{myenumlist}
All the experiments in the paper are carried out via simulations, which are fully reproducible and publicly available on GitHub, including the simulation software source code, the scripts to run the analysis, and the artifacts and plots.

The rest of this paper is structured as follows.
We summarize the system model assumptions and findings in~\cite{cicconetti_resource_2022} in \rsec{summary}.
%
%
We then review the related work on routing in quantum network in
\rsec{soa}.
The main contributions are reported in \rsec{eval-diff}, where we study the service differentiation, and in \rsec{eval-fair}, where we tackle the problem of fair sharing of resources.
\rsec{conclusions} concludes the paper and identifies the most important open research directions in this context.%


%
  \section{System Model}%
  \label{sec:summary}%
  In this section, we describe in short the quantum network abstract model adopted in the paper (\rsec{summary:network}), the resource allocation algorithm proposed in~\cite{cicconetti_resource_2022} (\rsec{summary:drr}), and the simulation methodology and tool (\rsec{summary:simulation}).
For more details, we refer the reader to \cite{cicconetti_resource_2022}, in particular sections II and IV, and references within.

\myssec{Quantum network model}{summary:network}

\myfigeps[scale=0.8]{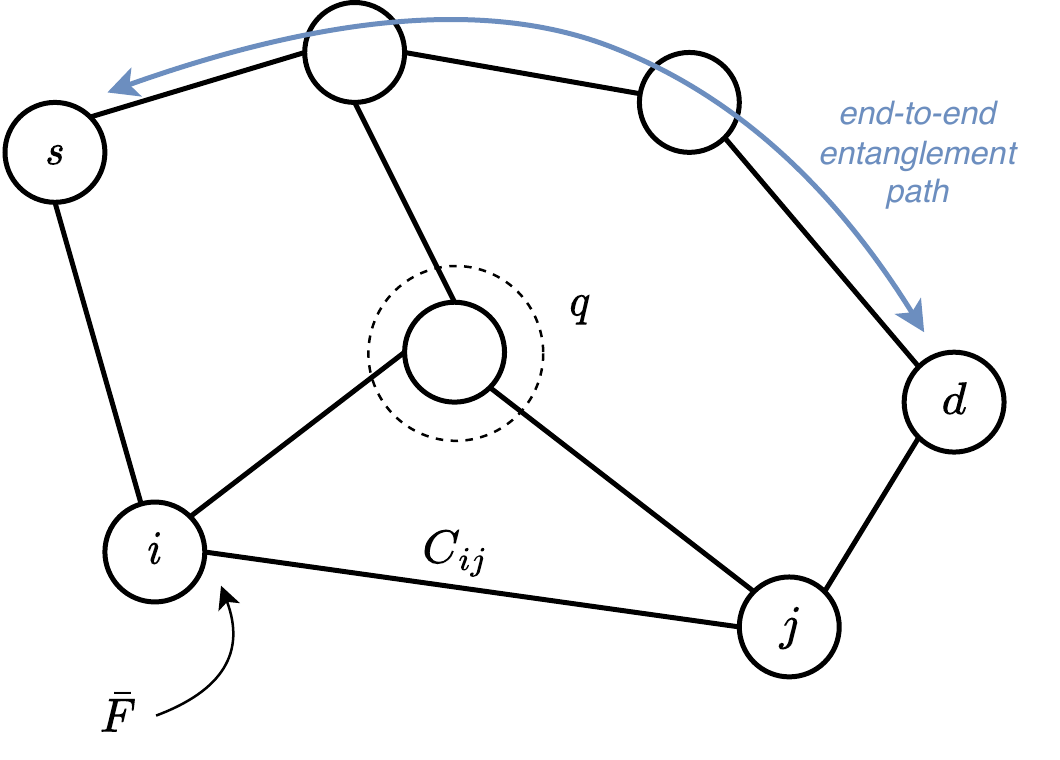}{Quantum network model. End-to-end entanglement can be established between two nodes $s$ and $d$ for which there is a path in $G(V,E)$, where the intermediate nodes perform entanglement swapping. $F$ is the fidelity with which the local link EPR pairs are generated; $q$ is the measurement success probability, which affects the entanglement swapping procedure; $C_{ij}$ is the capacity of edge $e_{ij}$, in EPR-pairs/s.}

The quantum network model is illustrated in \rfig{system-network-model} as a graph $G(V,E)$, where nodes represent quantum devices (repeaters or computers), and edges represent direct quantum communication links between them~\cite{chakraborty_entanglement_2020}.
We assume that maximally entangled EPR (Einstein–Podolsky–Rosen) pairs, e.g., $|\Phi^+\rangle$, are generated periodically at each link, and they have initial fidelity equal to $\bar{F} \in [0.5, 1]$.
The fidelity is a measure of how close a given quantum state is to a reference state, with 1 meaning that they are identical.
Every edge $e_{ij} \in E$ has a given capacity $C_{ij}$, which is the rate of generation of EPR-pairs, which in turn depends on the physical characteristics of the quantum network devices and links.
Quantum networks will offer the capability to produce entangled EPR-pairs between remote nodes, i.e., nodes that are interconnected through intermediate hops, which will perform the procedure of entanglement swapping to this purpose.
Such a procedure is stochastic in nature, in 1G quantum repeaters, and we assume that it succeeds with probability $q$~\cite{sangouard_quantum_2011}.
An end-to-end EPR-pair can only be used in a meaningful manner if all the entanglement swaps along the path have succeeded, which leads to the following formula to compute the maximum net rate that can be used by a quantum application consuming resources along a path $p = \{ (s,v_1), \ldots, (v_{N},d)\}$:

\begin{equation}\label{eq:net-rate}
    r(p) \leq \frac{\min_{e_{ij} \in p}\{C_{ij}\}}{q^{|p|-1}},
\end{equation}

\noindent where $|p|$ is the length of the path $p$, in number of edges. Furthermore, entanglement swapping reduces the fidelity of the end-to-end entangled EPR-pair according to the following formula~\cite{briegel_quantum_1998}:

\begin{equation}\label{eq:fidelity:perfect}
    F(p) = \frac{1}{4} + \frac{3}{4} \left( \frac{4\bar{F}-1}{3} \right)^{|p|}.
\end{equation}

\noindent We can say that $r(p)$ and $F(p)$ define the effective rate at which two end-points can transfer EPR-pairs, which is the logical equivalent of throughput in classical networks.

In our previous work~\cite{cicconetti_resource_2022} we have classified the quantum applications in two categories:

\begin{myitemlist}
    \item \textbf{Flows}. They are characterized by the need for two specific nodes to exchange a constant flow of EPR pairs in a point-to-point manner for the whole duration of a session. If the rate of EPR pairs falls below the requested amount, then the application \ac{QoS} degrades. Examples of such applications are: clock synchronization and \ac{QKD}.
    \item \textbf{Apps}. In this category we find distributed quantum computing applications, each characterized by a given quantum computer (\textit{host}) running an algorithm that pools the resources of a number of other quantum computers (\textit{peers} or \textit{workers}). There is no required EPR-pair rate, but the application wishes to consume as many EPR-pairs as possible to complete the execution faster. Such kind of service is equivalent to best-effort or elastic traffic in a classical data network. Since the network operator might provide a differentiated service, we foresee that each app is also assigned a \textit{weight} ($\rho$), which is a relative indication of how much throughput (in EPR-pair/s) it should be given in the long-term compared to another app with a different weight.
\end{myitemlist}

For both categories, we foresee the \textit{minimum fidelity} ($F^{\min}$) can be also a user requirement.
Since in this work we focus on distributed quantum computing applications, whose traffic is better modeled by apps than flows, in the rest of the paper we do not consider further the latter.

\myssec{QDRR resource allocation algorithm}{summary:drr}

We report below a recap of the apps' resource allocation algorithm in~\cite{cicconetti_resource_2022}, which in the following will be called Quantum DRR (QDRR) as it was inspired by the well-known \ac{DRR} algorithm~\cite{shreedhar_efficient_1995}.
%
%
The basic idea of QDRR is to provide all applications with a fair chance to be allocated a fraction of the quantum network resources.
This is enforced by visiting the applications in a round robin: at each visit, the application can be allocated capacity across multiple paths towards its peers, up until a given amount that is proportional to the application's weight.
Shortest paths are always preferred to longer ones, because they are more efficient.
A more detailed explanation follows.

The algorithm has two system parameters, which are set based on our previous results: $k = 4$ and the round size $\phi = 10$~EPR-pairs/s.
For a set of apps $i \in \mathcal{A}$, each defined by a host node $\{h_i\}$ and a set of candidate peers $\mathcal{W}_i$, QDRR consists of the following steps:

\begin{myenumlist}
    \setcounter{enumi}{-1}

    \item $\forall i \in \mathcal{A}$, $\forall j \in \mathcal{W}_i$:
    find $k$ shortest paths from $h_i$ to $w_{ij}$ and
    add them to $\mathcal{P}_i$; at the end, each $\mathcal{P}_i$ contains
    up to $k$ paths for each possible peer of $i$.
    Initialize the active list of apps $\mathcal{L}$ with the
    identifiers of all the apps $\mathcal{A}$.
    Copy the graph $G(V,E)$ into a temporary copy $G'$.
    
    \item  If $\mathcal{L} = \emptyset$ terminate.
    Otherwise, let $a$ be the next application to be visited in $\mathcal{L}$ in round robin order.

    \item Set the residual capacity that can be used by the current
    app $a$ in this round to $\delta_a = \phi \frac{\rho_a}{\sum_{i
    \in \mathcal{A}} \rho_i}$, i.e., a fraction of the round size $\phi$
    proportional to its priority.

    \item Select the shortest path $p \in \mathcal{P}_a$.
    The shortest path is the one that requires the least amount of resources
    among those available for the current app, according to \req{net-rate},
    \textit{and} gives the maximum fidelity, according to \req{fidelity:perfect}. \\
    If $p$ is not feasible anymore because it contains edges that have been
    removed from $G'$ discard it and move to the next
    shortest path.
    If $\mathcal{P}_a = \emptyset$ remove $a$ from the active
    list $\mathcal{L}$ and continue from Step~1.

    \item Determine the gross rate to be assigned to the current application $a$
    along path $p$ at this round as
    $R = \min\left\{\delta_a, \min_{e \in p} C_e\right\} \ge 0$.
    The corresponding net rate will be $r = R \cdot q^{|p|-1}$, as per
    \req{net-rate}.
    
    \item Remove $R$ from the capacity of all the edges along the path $p$.
    Remove from $G'$ all the vanishing edges.
    
    \item Update $\delta_a \leftarrow \delta_a - R$.
    If $\delta_a = 0$ restart from Step~1, otherwise continue from Step~3.
\end{myenumlist}

%

\myssec{Simulation methodology and tool}{summary:simulation}

We conclude the section by describing the methodology and tool adopted for the performance evaluation in \rsec{eval-diff} and \rsec{eval-fair}.

Like in~\cite{dai_optimal_2020}, we use a \ac{PPP} to generate the position of an average of $\mu$ nodes in a flat square grid with edge size 60~km; a link is added between two nodes with probability $p_{\mathrm{link}} = 0.5$ if their Euclidean distance is smaller than a threshold $\tau$.
The capacity of each link is drawn from a r.v. uniformly distributed
between 1~Bell pair/s and 400~Bell pairs/s, as
in~\cite{chakraborty_entanglement_2020}.
The initial fidelity of Bell pairs is $\bar{F} = 0.95$, which is widely used in the literature, and the entanglement swapping success probability is $q = 0.5$, which is the best value that can be obtained with linear optics components~\cite{sangouard_quantum_2011}.
Based on previous results in~\cite{cicconetti_resource_2022} we have selected two representative topologies:

\begin{myitemlist}
    \item \textit{dense}: $\mu = 100$, $\tau = 20$~km;
    \item \textit{sparse}: $\mu = 50$, $\tau = 15$~km.
\end{myitemlist}

\noindent The following metrics are used to evaluate the performance:

\begin{myitemlist}
    \item The \textit{net rate} of the apps, i.e., the number of EPR-pairs that the end-points can consume in the unit of time, which is a direct operational measure from the point of view of the end users;
    \item The \textit{max-min fairness}, which is the difference between the top and lowest net rates assigned to the apps.
    \item The \textit{fidelity}, weighted for each app on the net rate assigned to the corresponding peer, which impacts on the accuracy and convergence of the distributed \ac{QC} applications.
    \item The \textit{inter-class unfairness index}, for a class of apps with $R$ priority weights $\rho_j$, provided that each app is allocated net rate $r_i$, defined as:
\begin{equation}\label{eq:unfairness}
    \sqrt{\sum_{i=2}^{R} \left( \frac{r_i}{r_1}  - \frac{\rho_i}{\rho_1} \right)^2 }
\end{equation}
    This measures the distance of the net allocations with respect to an ideal case where the proportions between rates is exactly the same as the proportions between priorities.
    %
    %
\end{myitemlist}

We used a Monte Carlo approach: for any combination of the parameters
under study, we simulated 6,000 drops with randomly generated networks and random workload.
Statistical significance has been verified for all the metrics in
the experiments performed, but we seldom include error bars in plots
for better readability.
The simulation tool used is a custom simulator, developed in C++ and using the Boost Graph Library, available as open source under a MIT license on GitHub:

\noindent\begin{center}
    \url{https://github.com/ccicconetti/quantum-routing/}
\end{center}

For full reproducibility, the repository also includes the scripts to run the experiments, as well as the artifacts obtained and the Gnuplot files to produce the plots: see tag \texttt{v1.5}, experiments labeled \texttt{004} and \texttt{005}.%

  \section{Related Work}%
  \label{sec:soa}%
  The literature on quantum networking and distributed \ac{QC} is not vast:
even though the basic ingredients
have been known since a long time ago ---consider for instance the
seminal paper by Bouwmeester~\textit{et al.} on quantum
teleportation~\cite{bouwmeester_experimental_1997} published on Nature
in 1997--- only recently there have been investments in an order of
magnitude sufficient for technology to take off.
This revamped interest has triggered new research
activities in this area, briefly reviewed below.

In general terms, the problem of \textbf{quantum routing} is formulated
as follows: given a network of quantum nodes (repeaters or computers) and
a set of traffic flows identified by their sources, destinations, and
application requirements (e.g., the minimum fidelity),
find the ``best'' paths that fulfill the constraints.
Some works have studied the problem by reusing the findings in the
area of routing in classical networks.
Van Meter~\textit{et al.} proposed a quantum version of the famous Dijkstra's
shortest path algorithm, which was shown to give very
good performance with an appropriate selection of the routing
metric that considers the specific properties of quantum
networks~\cite{van_meter_path_2013}.
More recently, Caleffi~\textit{et al.} have proposed a slightly less efficient
variation of Dijkstra's algorithm that can work with non-isotonic routing
metrics, which they have advocated to provide superior performance in
selected use cases~\cite{caleffi_optimal_2017}.
Dijkstra's algorithm is also the subject of~\cite{chakraborty_distributed_2019},
where the authors lay some mathematical foundations that allow them to
derive upper bounds of performance in specific network topologies, including
grid and ring.

A different direction is explored by Pant~\textit{et al.}, who
studied the distribution of routing information to the
nodes~\cite{pant_routing_2019}; for this they propose a time-slotted
approach: in the first part of the slot every repeater tries to
create a local entanglement with all its neighbors, then in the
second part the paths are established as instructed by a centralized
authority.
One interesting aspect of the paper is that multiple paths are selected
for the same (source, destination) to maximize the rate of end-to-end EPR-pairs.
We have also adopted this time-slotted model in \cite{cicconetti_request_2021},
where we have investigated the issue of ``scheduling'' of traffic flows, i.e.,
determining the order in which to assign paths to pending requests, in case
the network resources are not sufficient to serve them all.
This problem is called ``distribution'' in~\cite{dai_optimal_2020}, where
the authors formulate it as an \ac{ILP}, for which they derive closed
formula performance bounds in the case of a homogeneous chain of
quantum repeaters.
The issue is also addressed in~\cite{chakraborty_entanglement_2020},
where the authors have proposed to split the overall quantum routing
problem in two to reduce the computational complexity: first, they
determine the rates achievable by the traffic flows under the given
network constraints using an approach based on multi-commodity flow
optimization, then they map these rates to paths.
The paper adopts a network model using probabilistic entanglement swapping,
which we reuse in this work (described in \rsec{summary}).

An important reference for our study is~\cite{li_effective_2021}, where
the authors study the allocation strategy of traffic flows for which
the paths have been pre-determined: they do so by borrowing the fairness
concept from data networks and re-using traditional algorithms from
the relevant literature.
In our paper, we also borrow from the same literature, though we apply
the concepts to a different class of applications, as it will be clear
in the next section.
\textit{As a matter of fact, all the scientific works cited above
have focused on point-to-point traffic flows, while in this paper
we focus on a different type of traffic that is more suitable to
model distributed \ac{QC}, with distinguishing features that do not
allow the reuse of state-of-the-art solutions.}
Rather, we claim that any existing routing/allocation/scheduling
solutions should work \textit{in parallel} to our proposed scheme
to provide an effective resource allocation to each of the two
traffic classes.

In addition to mere routing aspects, system-wide studies have also been
published.
We mention~\cite{van_meter_quantum_2021}, which is a compendium of
several previous studies from the same authors that illustrates an
overall architecture of the Quantum Internet, also including
application, protocol, and deployment aspects at a high level.
On the other hand, other works have focused on specific components,
which are complementary to the research activity presented, e.g.,
\cite{zhao_quantum_2021} on congestion control in transport protocols
and~\cite{dahlberg_link_2019} on the link layer, with a focus on
hardware and physical-layer considerations.
%

Furthermore, some research groups have been working to define the
basic principles of \textbf{distributed \ac{QC}}.
Parekh~\textit{et al.} have defined an elegant framework for the parallel
execution of a broad class of quantum algorithms on multiple
nodes~\cite{parekh_quantum_2021}, both using remote entanglement and
with \ac{LOCC} only, also studying in depth three classes of algorithms:
variational quantum eigensolver, low-depth quantum amplitude estimation,
and quantum k-means clustering.
In~\cite{cuomo_optimized_2021} the authors address the problem of
the efficient compilation of circuits for distributed \ac{QC}
by considering that some gate operations will be executed remotely, hence
with much different latency and reliability than on-chip operations.
The research of Dahlber~\textit{et al.} moved in the same direction
and went as far as defining a set of low-level instructions (called
NetQASM) for distributed \ac{QC} systems seamlessly
supporting local and remote gates~\cite{dahlberg_netqasmlow-level_2022}.
%
\textit{These works confirm that there is a growing interest in
distributed \ac{QC}, which is a motivation for our work.}
%

On another line of research, solutions have been proposed to trade capacity for fidelity, by using \textbf{purification} (or distillation) techniques~\cite{van_meter_system_2009}.
In brief, they consist in entangling multiple pairs of qubits with low fidelity and then collapsing them into a single one with high fidelity.
We do not consider network-level purification in this work to remain consistent with the positioning of our contribution within the realm of 1G-repeater quantum networks.
End-to-end purification is also possible, that is the operation is performed by quantum computers after the qubits have been entangled all along the path(s).
This is studied, e.g., in~\cite{zhao_e2e_2022}, where the authors propose a quantum routing algorithm that maximizes the rate of EPR-pairs, while deciding not only the paths but also the purification patterns.
\textit{These works complement our contributions since they operate on constant-rate point-to-point flows only and they do not take into account network provisioning issues.}

Finally, in line with the vast majority of prior works, we only consider bi-partite entanglement, i.e., made of two qubits, each situated in a quantum computer.
While there are some promising theoretical studies on repeater-assisted multi-partite entanglement, i.e., involving more than two qubits (e.g., \cite{pompili_realization_2021}), the research in that area is in still its infancy.
One noteworthy contribution is~\cite{patil_entanglement_2022}, where the authors propose to adopt $n$-fusion of bi-partite entanglements to create higher level entanglements between $n > 2$ quantum computers.
An appealing property that they demonstrate, under some assumptions, is that the entanglement rate between nodes remains constant with increasing distance, in number of hops.
Ways to exploit this phenomenal quality are still under study.
\textit{Multi-partite entanglement in a quantum network is generated starting from elementary bi-partite entanglements, which is the subject of this work.}
%



%
  \section{Service Differentiation}%
  \label{sec:eval-diff}%
  In this section we extend the study in~\cite{cicconetti_resource_2022} by analyzing the QDRR algorithm along two directions which have remained so far uninvestigated: service differentiation by assigning apps different $\rho$ values (\rsec{eval-diff:prio}) and apps with different fidelity thresholds $F^{\min}$ (\rsec{eval-diff:fidelity}).
In all the simulations in this section the peers are selected as follows: for each app $i$ on node $v$ we draw at random between 2 and 4 candidate nodes by sampling in a uniform manner from the set of all nodes that are reachable from $v$ in 2--7 hops.
For benchmarking purposes, QDRR is compared to two baseline algorithms: random and best-fit.
The Step~0 in \rsec{summary:drr} is the same for all the algorithms, that is for each app we find $k = 4$ shortest paths to reach any peer.
All the algorithms then loop through all the possible paths for all candidates for each app until there are no more feasible paths to be assigned, but they differ in how they do so: QDRR is descriped by Steps~1--6 in \rsec{summary:drr} (and in far more details in Sec.~IV-C in~\cite{cicconetti_resource_2022}), while:

\begin{myitemlist}
    \item \textbf{Random}: at each iteration one app with remaining paths is chosen at random and assigned its shortest path among any of its peers, which is allocated the maximum rate along the path.
    Random is representative of allocation algorithms that provide fair access to the quantum network resources in a per-app manner.
    \item \textbf{Best-fit}: at each iteration, select the app with the shortest path to reach one of its peers and allocate the maximum rate along that path.
    Best-fit is representative of allocation algorithms that strive to maximize the efficiency, that is the ratio between net entanglement rate and the quantum network capacity allocated.
\end{myitemlist}

\myfigeps{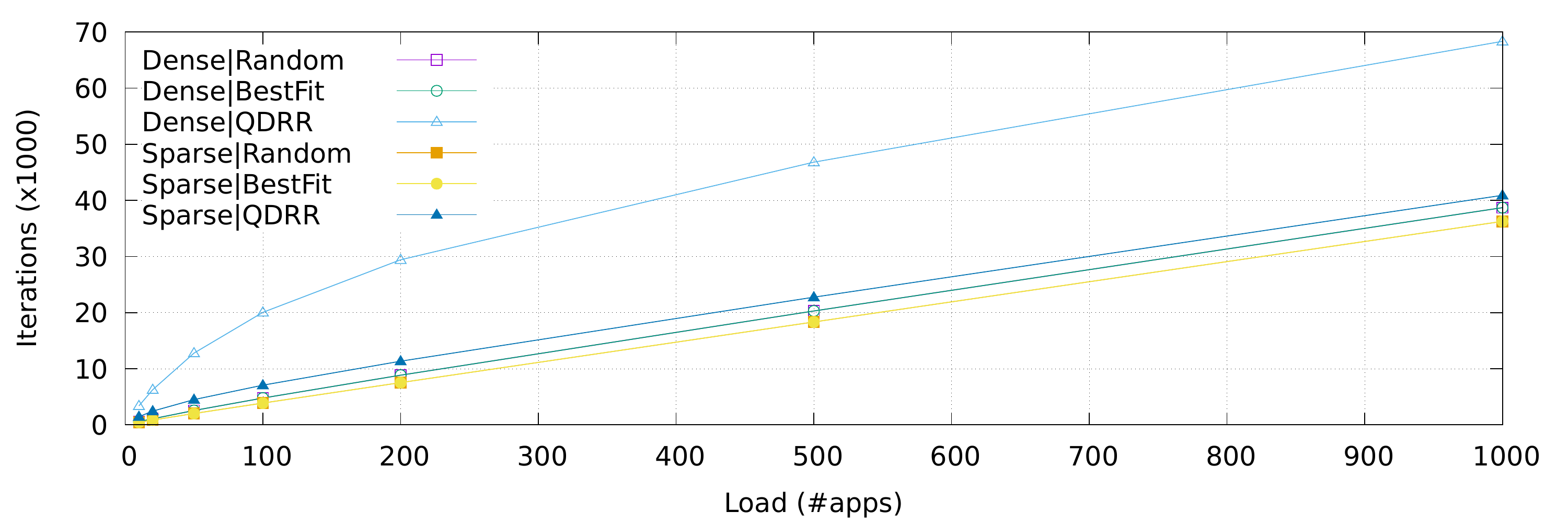}{Number of iterations (expect Step~0 in \rsec{summary:drr}) with random, best-fit, QDRR, in a dense vs.\ sparse topology, when increasing the number of apps with $\rho \in \{1,2,4\}$.}

Unlike QDRR, both random and best-fit can be considered greedy algorithms, since they never backtrack to a previously selected combination of (app, peer, path), which is always allocated as much throughput as possible.
Therefore, their worst-case computational complexity (expect Step~0) is $O\left( k |\mathcal{A}| \log|\mathcal{A}| \mathbb{E}[\mathcal{W}_i] \right)$: $k$ is the number of shortest paths selected in Step~0; $|\mathcal{A}|$ is the number of apps;  $\log|\mathcal{A}|$ takes into account the random selection or the extraction from an sorted data structure, respectively for the random and best-fit resource allocation algorithms; and, $\mathbb{E}[\mathcal{W}_i]$ is the average number of peers per app.
The complexity of QDRR is discussed in~\cite{cicconetti_resource_2022} and it depends on the choice of $\phi$.
To give an idea of the relative average complexity between QDRR and random/best-fit, we report their number of iterations in the simulations discussed in \rsec{eval-diff:prio} below in \rfig{004-mp-visits}.
As can be seen, in a sparse scenario the time complexity of QDRR is only marginally higher than that of greedy algorithms random and best-fit, but it becomes clearly higher in a dense scenario.
If this is an issue, the value of $\phi$ can always be tuned to reduce the number of iterations, trading off fairness for speed.

\myssec{Different traffic priorities}{eval-diff:prio}

\myfigeps{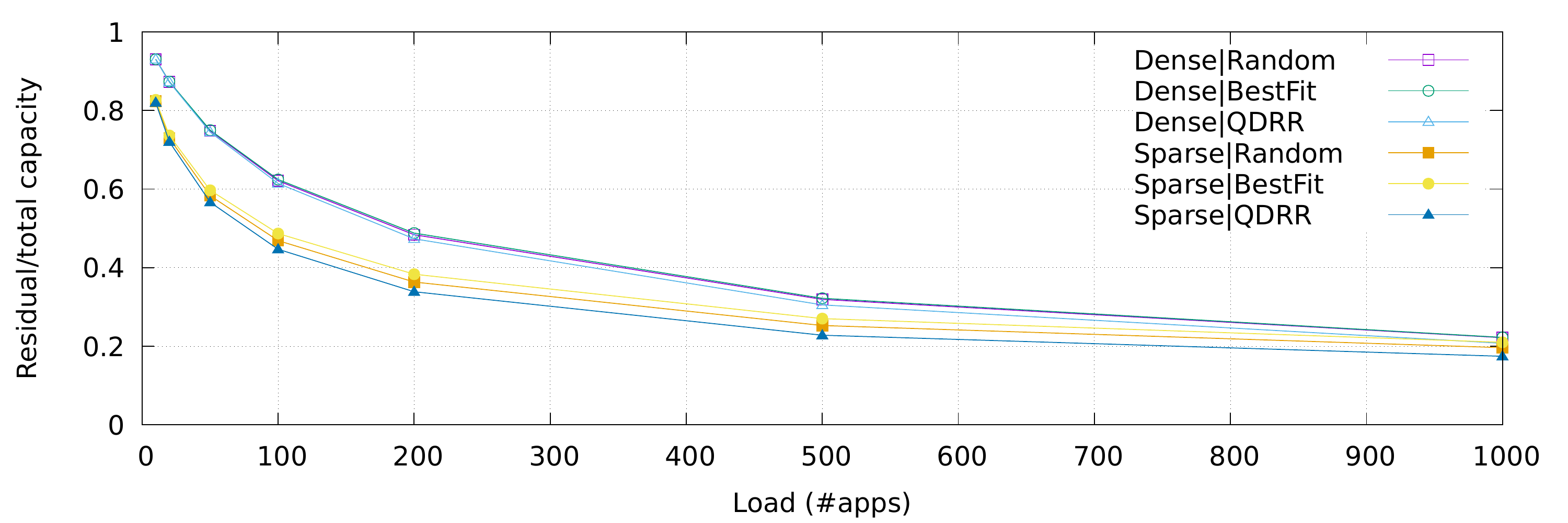}{Ratio between the residual capacity and the total capacity with random, best-fit, QDRR, in a dense vs.\ sparse topology, when increasing the number of apps with $\rho \in \{1,2,4\}$.}

In a first batch of results we increase the load of the network from 10 to 1000 apps.
For every app, its priority weight $\rho$ is drawn randomly in $\{1,2,4\}$, while $F^{\min}$ is the same for all and equal to 0.7.
As shown in \rfig{004-mp-residual}, for all the allocation algorithms and topologies the relative residual capacity decreases with a sub-linear trend as the load increases, which is due to the exponential relation between the net rate, in EPR-pairs/s, and the number of hops as per \req{net-rate}.
The utilization is higher in a sparse topology, while the difference between the allocation algorithms is negligible.
In the following we report only the results in a dense topology, due to limited space; the complete results can be retrieved from the public GitHub repo above.



\myfigeps{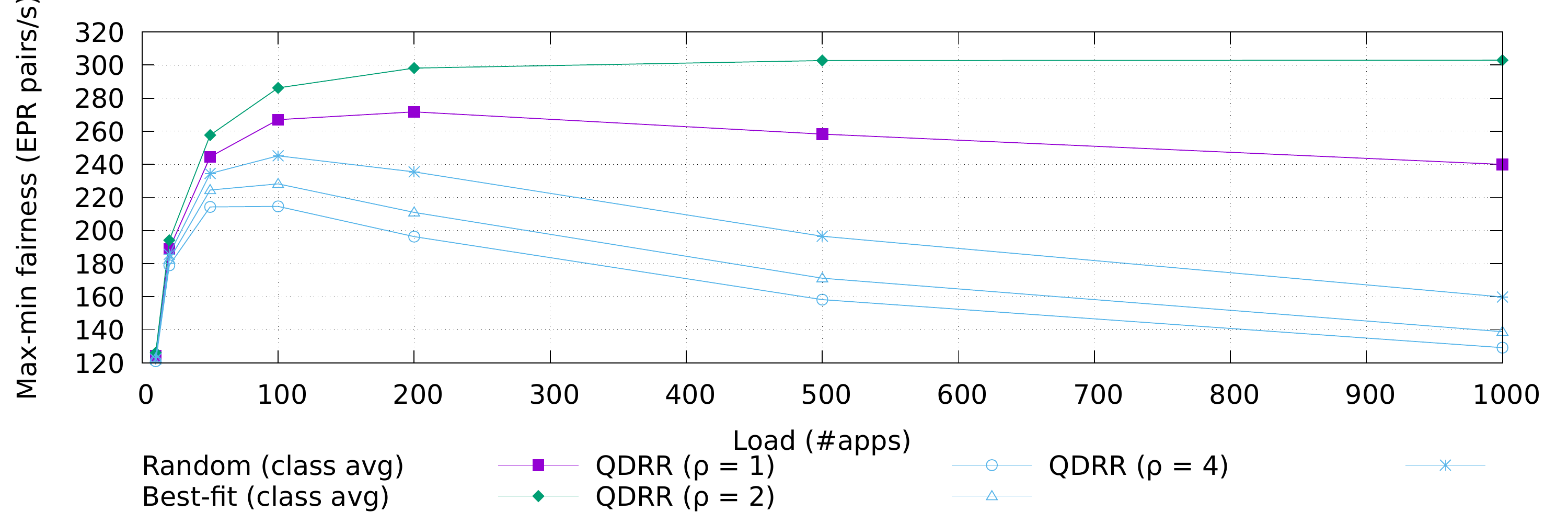}{Max-min fairness with random, best-fit, QDRR, in a dense topology, when increasing the number of apps with $\rho \in \{1,2,4\}$.}

We begin by showing the max-min fairness in \rfig{004-mp-jitter-dense}. 
Since random and best-fit do not differentiate based on the $\rho$ values, for them we show an aggregate average, while we keep separate curves for QDRR.
With very low loads, the max-min fairness is good, i.e., low, for all allocation algorithms, because there is little contention on resources.
However, as the load increases, the max-min fairness increases steeply and the behavior is significantly affected by the allocation algorithm: with random the curve reaches a peak, which then slowly decreases towards high loads; best-fit performs worst, as expected, by continuing to increase, even though only slightly after the initial spur; for all traffic categories, QDRR provides increasingly better fairness as the load increases.
The latter can be explained as follows.
When there are few apps, it is likely that there are not many shared nodes/paths, hence QDRR does not really have a chance to distribute the resources proportional to the apps' weights; on the other hand, with more apps, it is increasingly easier for QDRR to enforce priorities by regulating the resources in common.

\myfigeps{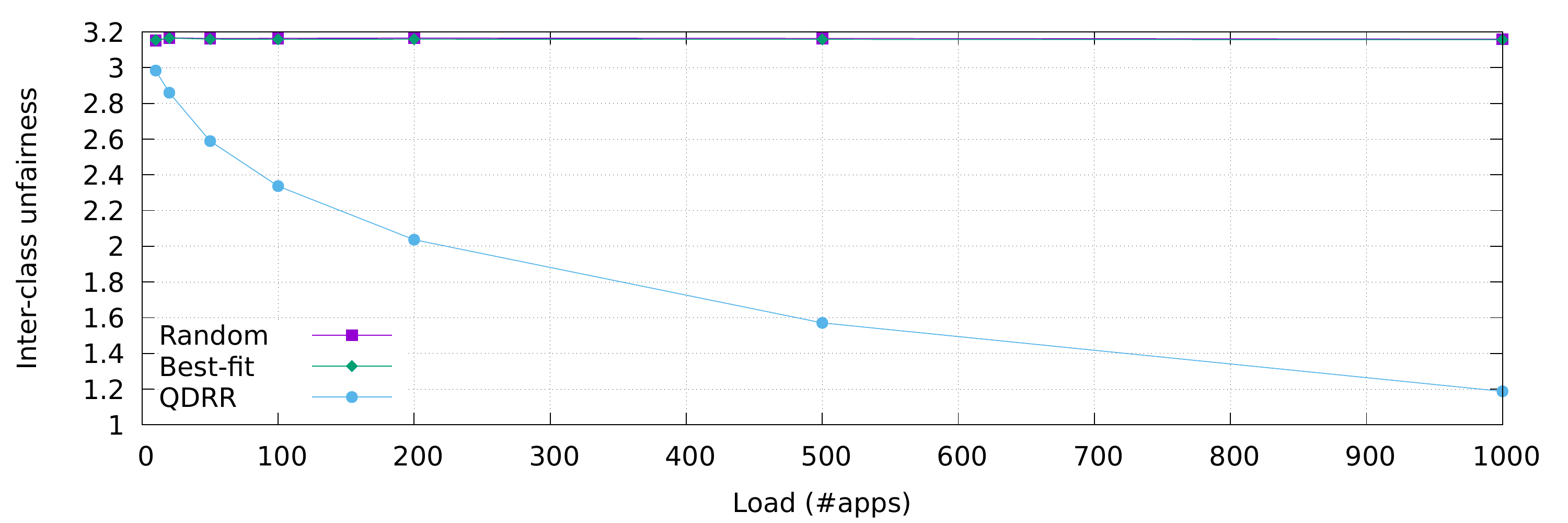}{Inter-class unfairness index with random, best-fit, QDRR, in a dense topology, when increasing the number of apps with $\rho \in \{1,2,4\}$.}

To better show the service differentiating behavior of QDRR, we show the inter-class unfairness index, as defined in \req{unfairness}, in \rfig{004-mp-netrate-dense-ndx}.
It is clear that QDRR is the only allocation algorithm providing the apps with a clear service differentiation, which improves as the load increases for the same reason above.

\myfigeps{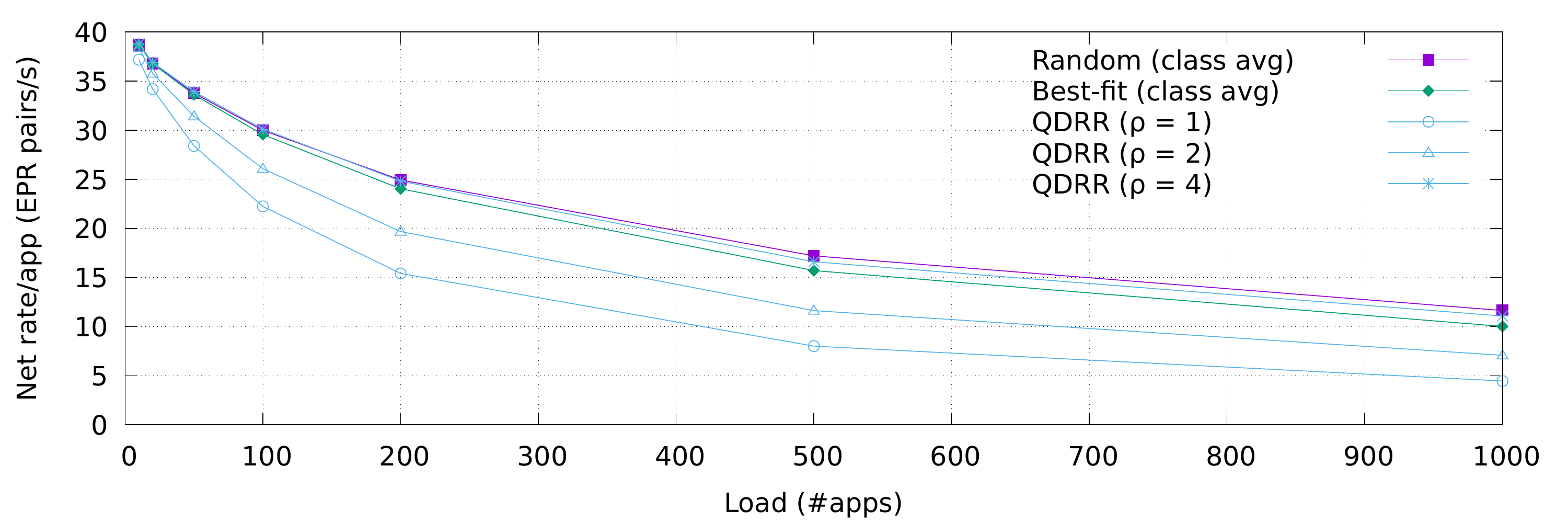}{Net rate/app with random, best-fit, QDRR, in a dense topology, when increasing the number of apps with $\rho \in \{1,2,4\}$.}

Finally, in \rfig{004-mp-netrate-dense-app} we show the net rate/app, in EPR-pairs/s: even though it slowly decreases for all the resource allocation algorithms, we can see that random and best-fit achieve better rates.
Therefore, \textit{QDRR is effective in differentiating service across apps with different priority weights, but this incurs a cost, in terms of net rate.}

\myssec{Different fidelity thresholds}{eval-diff:fidelity}

\myfigeps{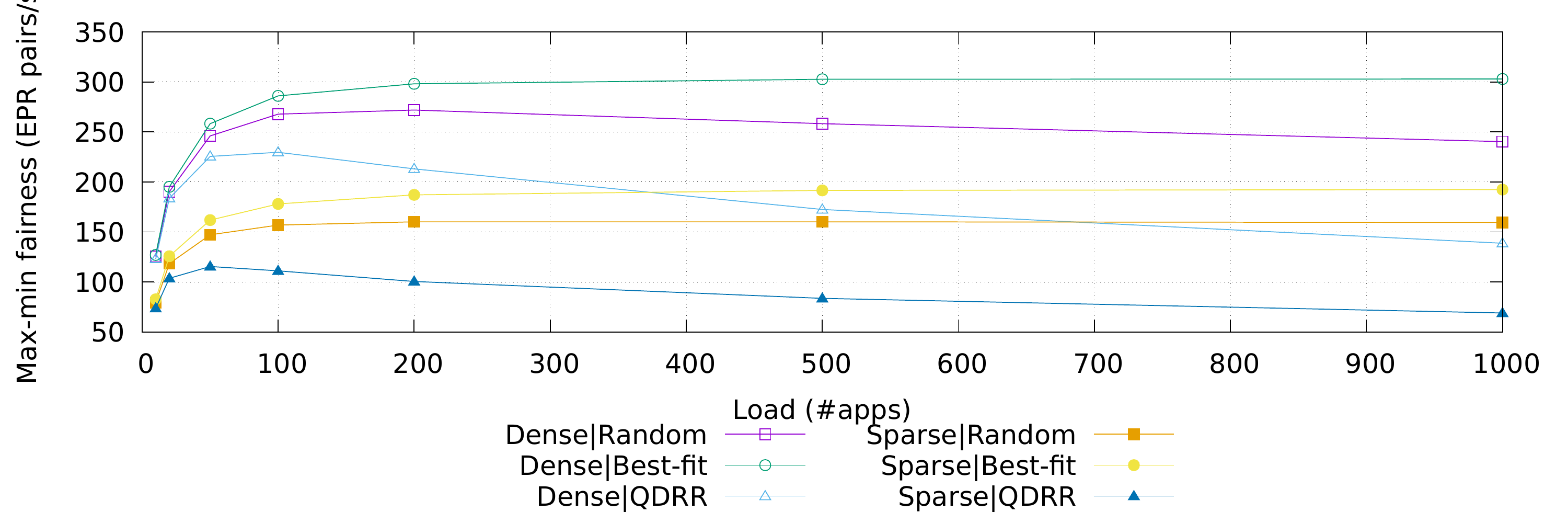}{Max-min fairness with random, best-fit, QDRR, in dense vs.\ sparse topologies, when increasing the number of apps with $F^{\min} \in \{ 0.7, 0.8, 0.9\}$.}


We now report the results obtained in a new batch, which follows the same direction as above, but we set $\rho=1$ for all apps and draw randomly the minimum fidelity $F^{\min}$ from $\{ 0.7, 0.8, 0.9\}$, instead.
We show the max-min fairness in \rfig{004-mf-jitter-app}, for both topologies.
Like in \rsec{eval-diff:prio}, QDRR achieves significantly better performance than random and best-fit, the latter performing worst.

\myfigeps{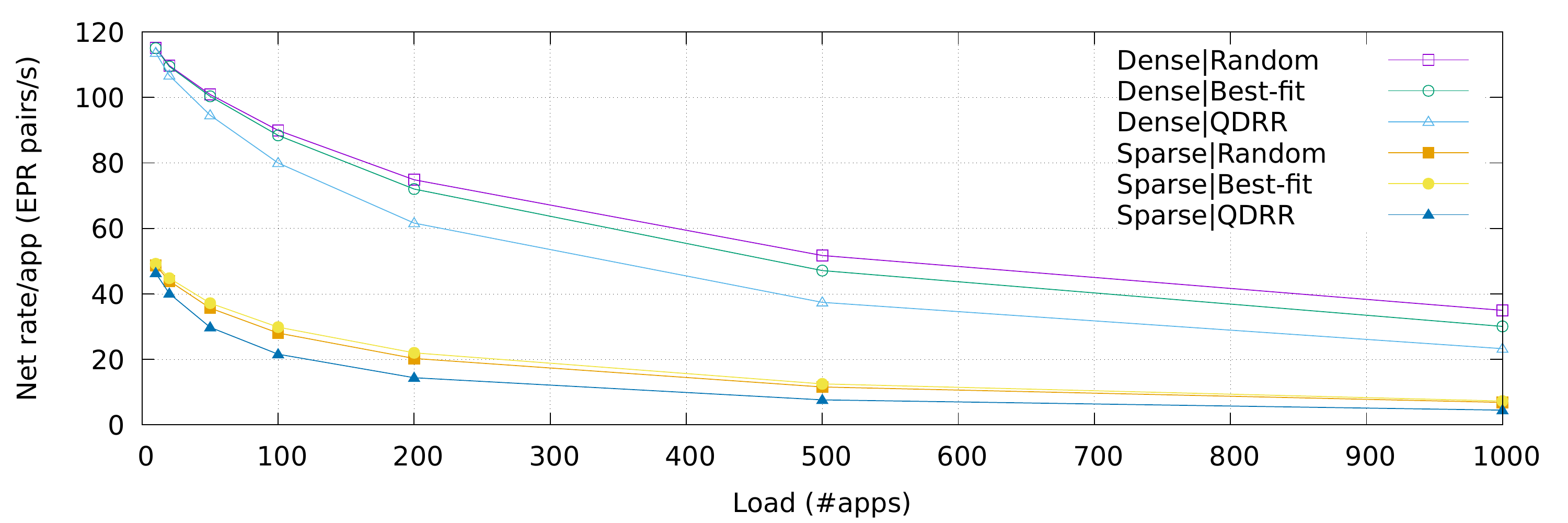}{Net rate/app with random, best-fit, QDRR, in dense vs.\ sparse topologies, when increasing the number of apps with $F^{\min} \in \{ 0.7, 0.8, 0.9\}$.}

However, as can be seen in \rfig{004-mf-netrate-app}, also with a mix of fidelity thresholds, the net rate/app of QDRR is slightly less than that with both greedy allocation strategies, which confirms the trade-off already identified in the previous batch of experiments.
It is worth noting that such a trade-off is well-known also in different contexts: for instance, in cellular systems, greedy scheduling algorithms that prioritize user terminals with good channel conditions (often known as ``max C/I'') are bound to provide a higher cell throughput at the cost of an inferior fairness compared to milder strategies such as Proportional Fair~\cite{jalali_data_2000}.

\textit{In conclusion, like for heterogeneous weights, QDRR can provide service differentiation to apps with different fidelity thresholds, but the net rate achievable is slightly reduced compared to alternatives that do not differentiate.}%

  \section{Fair Sharing}%
  \label{sec:eval-fair}%
  In this section we address a problem that is preliminary and complementary to that defined in our previous work~\cite{cicconetti_resource_2022} and investigated in \rsec{eval-diff} with differentiated service.
So far we have assumed that all nodes are equal, and each host is wishing to cooperate with a given set of workers, without elaborating further on how such a set is selected; for performance evaluation purposes, such a set was selected randomly from candidates depending only on the distance as per the specific scenario simulated.
In the following, instead, we address specifically this issue: indeed, \textit{how does one decide which are the possible workers of a host node?}

\myfigeps{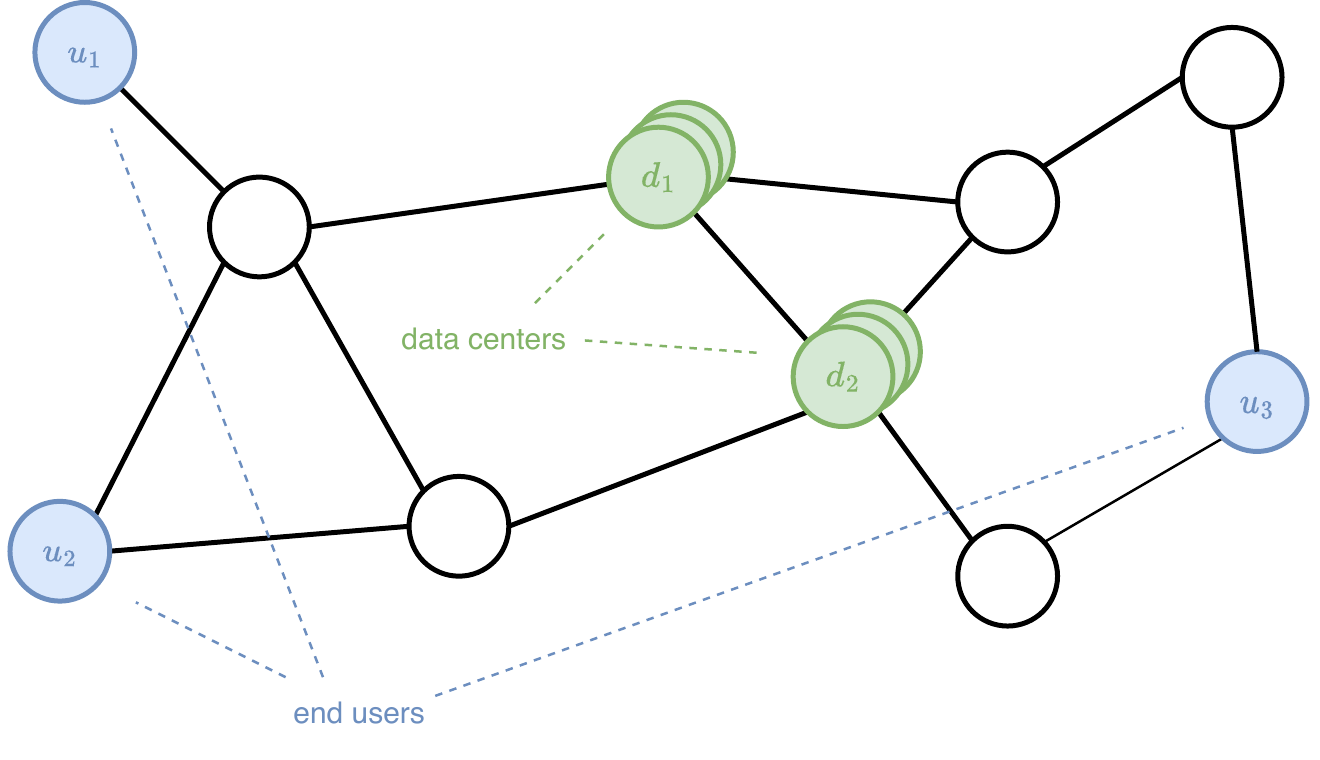}{Example of quantum network with three end users, labeled from $u_1$ to $u_3$, wishing to host distributed quantum computing algorithms with data center nodes $d_1$ or $d_2$\added{, represented with multiple co-located circles to indicate that they are expected to be more powerful than end users and, possibly, they might have a more complex internal structure that is not elaborated further in this paper}; the other nodes in $G(V,E)$ participate to the end-to-end entanglement of qubits as intermediate hops.
Like in \rsec{summary:network}/\rfig{system-network-model}, the network is characterized by capacity $C_{ij}$ of the link between nodes $i$ and $j$, initial generation fidelity $F$, and entanglement swapping success probability $q$.}

We adapt our system model to the new landscape by \added{specializing the role of nodes}.
As illustrated in the example in \rfig{system-fair-sharing-model}, we assume that nodes can be of three types:
\begin{myinlinelist}
    \item \textit{end users} ($u_i$): \added{they act as the ``home \acp{QC}'' of customers wishing to run quantum algorithms on them, also exploiting quantum computation resources offered by other nodes through distributed quantum computing; a customer operates the end user via a classical computer for, e.g., input preparation, circuit compilation, and quantum network resource reservations;}
    \item \textit{data centers} ($d_j$): these are \acp{QC} that can \added{provide customers with extra quantum computation capacity to be added to their respective end users} for the purpose of solving bigger instances of their problems via distributed quantum computing, exploiting end-to-end entanglement of qubits through an underlying quantum network;
    \item \textit{intermediate nodes}: quantum repeaters who do not consume or offer \ac{QC} capacity but contribute to the quantum network by performing entanglement swapping between links as instructed by the resource allocation algorithm (e.g., QDRR).
\end{myinlinelist}
%
%
\added{A node can play any combination of the three roles above}.
We assume that end user $u_i$ may perform distributed quantum computing with any combination of data centers $\{ d_j \}$\added{, following commercial agreements that are out of the scope of this work}.
All other quantum network assumptions in \rsec{summary:network} remain the same.
In \rsec{eval-fair:assignment} we formulate the problem in mathematical terms and propose a solution, which is then evaluated via simulation in \rsec{eval-fair:eval}, compared to two alternatives.

\myssec{\acl{QWAP}}{eval-fair:assignment}

In its most general formulation, the problem of how to best assign each host a set of workers depends on several factors that may be not known or under the control of the quantum network operator.
They include, for instance: the quantum algorithms to be run, the different characteristics of the end user and data center \acp{QC}, the schedule of the task execution, not to mention administrative factors (contracts, partnerships, billing issues) and technical constraints (do the \acp{QC} need to have the same hardware or software?).
Since both quantum networking and distributed quantum computing are in their infancy, we consider unrealistic to address the problem under such general settings.
Rather, we focus on aspects that are captured by our network model (in \rsec{summary:network}) with the goal of providing an initial understanding of the problem, to be used as a stepping stone by future studies as the technologies involved become more mature.
Our formulation, in natural language, is the following:

\noindent \underline{\ac{QWAP}}: find the sets of workers, selected from the data center nodes, to be assigned to each host, from the end user nodes, so that the overall profit of the hosts is maximum while balancing the load of data centers.

The problem can be formulated in a formal manner once we define the notions of ``profit'' and ``load balancing''.
Based on the prior works in the literature (see \rsec{soa}), we consider the net entanglement rate in \req{net-rate} as the \textbf{profit}, i.e., between two possible data centers $d_1$ and $d_2$ considered as candidate workers for end user $u$, we prefer the one that potentially achieves the highest net entanglement rate.
On the other hand, we introduce \textbf{load balancing} as follows.
First, we define the system parameter $W$ as the target number of workers per end user.
Then, we force each data center to be assigned as worker to at most $B$ end users, where $B$ is determined as the minimum value that allows this constraint to be provided with given $W$, $N_u$ end users, and $N_d$ data centers.
This way, we force an even load across data centers.
Note that this formulation can be trivially extended to the case where data centers have different capabilities by introducing appropriate weights, which we do not consider in this work to keep the notation succinct.

Assuming without loss of generality, again to simplify notation, that all end users have one and only one request to run distributed quantum computing, the \ac{QWAP} can be expressed as an optimization problem with objective function:

\begin{equation}\label{eq:wap-obj}
    \max \sum_{u=1}^{N_u} \sum_{d=1}^{N_d} \pi_{ud} x_{ud}
\end{equation}

\noindent such that:

\begin{align}
    \sum_{u=1}^{N_u} x_{ud} \leq B & & d=1,\ldots,N_d \label{eq:wap-c1} \\
    \sum_{d=1}^{N_d} x_{ud} \leq W & & u=1,\ldots,N_u \label{eq:wap-c2} \\
    x_{ud} \in \{0,1\} \label{eq:wap-c3} \\
    B=\left\lceil \frac{N_u \cdot W}{N_d} \right\rceil \label{eq:wap-c4} 
\end{align}

\noindent where the profit $\pi_{ud}$ between end user $u$ and data center $d$ is defined by:

\begin{equation}\label{eq:profit}
    \pi_{ud} =
      \begin{cases}
        0 & \text{if}~\forall p \in \mathcal{P}_{u,d}: F(p) \leq F_u^{\min} \\
        r_{ud\bar{p}} & \text{where}~\bar{p}=\arg\max_{p}\left\{ r_{udp}|F(p)\geq F_u^{\min} \right\} \\
      \end{cases},
\end{equation}

\noindent where $F(p)$ is the fidelity of the end-to-end entanglement along the path $p$, according to \req{fidelity:perfect}, $F_u^{\min}$ is the minimum fidelity requested by end user $u$, $\mathcal{P}_{u,d}$ is the set of all paths between $u$ and $d$ in $G(V,E)$, and the net rate $r_{udp}$ between end user $u$ and data center $d$ along path $p$ is as follows:

\begin{equation}\label{eq:rate-udp}
    r_{udp} = \frac{\max_{(i,j)\in p}\left\{ C_{ij} \right\} }{q^{|p|-1}}.
\end{equation}

The output assignment integer variables, as per \req{wap-c4}, are the $x_{ud}$, with the constraints as follows:
\req{wap-c1} ensures that no data center is assigned more than its fair share of $B$ users, where $B$ is computed via \req{wap-c4};
\req{wap-c2} ensures that no end user is assigned more than $W$ workers.
The profit $\pi_{ud}$, defined through Eqs.~(\ref{eq:profit}--\ref{eq:rate-udp}), corresponds to the maximum net rate of EPR-pairs/s that can assigned to end user $u$ along any path towards data center $d$ that fulfils its minimum fidelity requirement.
Before proceeding, we state two key observations about the \ac{QWAP}.

\noindent\underline{Observation\#1}. In a general graph $G(V,E)$, the number of paths between two nodes can be exponential with the size of the graph. For instance, in a complete graph this number is $\left\lfloor(V-2)!e\right\rfloor$. Therefore, the preparation of the problem input in \req{profit} might be very computation-intensive in practice.
In the evaluation below, we adopt a reasonable approximation: rather than finding all the paths $\mathcal{P}_{ud}$ between nodes $u$ and $d$, we use a reduced set $\mathcal{P}^{\bar{k}}_{ud}$ that consists of the $\bar{k}$ shortest paths ($\bar{k} = 10$ in the simulations in \rsec{eval-fair:eval}).
The rationale is that long paths have a small chance of being selected as $\bar{p}$ in the second branch of \req{profit}, because the net rate decreases exponentially with the path length as per \req{rate-udp}.

\noindent\underline{Observation\#2}. When $W = 1$, then Eqs.~(\ref{eq:wap-obj}--\ref{eq:rate-udp}) above can be trivially transformed into an assignment problem, whose optimum solution can be found efficiently.
With $W > 1$, however, the \ac{QWAP} becomes a ``multiple knapsack problem'', which is NP-hard, but for which several efficient heuristics are well-known in the operations research literature (e.g.,~\cite{martello_bound_1981}).

Based on the two observations above, we propose our \textbf{load balancing} algorithm to solve the \ac{QWAP}, which we implemented in our simulator and evaluated in the next section.
The idea of the load balancing algorithm is to find the optimal allocation for the first worker of each end user using the Hungarian algorithm~\cite{kuhn_hungarian_1955}, which finds the best (exact) solution in assignment problem instances; then, it progresses by considering one more worker at a time, until $W$, each time invoking the Hungarian algorithm again on the data centers with residual slots.
The algorithm is \textit{greedy} because it never backtracks prior decisions and it always terminates after a fixed number of iterations.
More formally, the load balancing algorithm consists of the following three steps:

\begin{myenumlist}
    \item Prepare the problem input, in particular the profits $\pi_{ud}$, by finding up to $\bar{k}$ shortest paths between $u$ and $d$ in $G(E,V)$ using Yen's algorithm~\cite{yen_finding_1971}.
    \item Determine $B$ based on $N_u$, $N_d$, and $W$ using \req{wap-c4}.
    \item Arrange the output of Step~1 in a profit matrix where the rows are the end users and there are $B$ columns for each data center, i.e., the matrix size is $N_u \times B N_d$.
    Then run $W$ iterations of the Hungarian algorithm.
    After each iteration, set $\pi_{ud} \leftarrow 0$ in all columns where a data center has been assigned (avoids that the constraint \req{wap-c1} is violated) and in all cells that refer to the same pair $u,d$ that has been assigned (avoids that a user is assigned multiple times the same data center).
\end{myenumlist}

\myssec{Evaluation}{eval-fair:eval}

\myfigeps{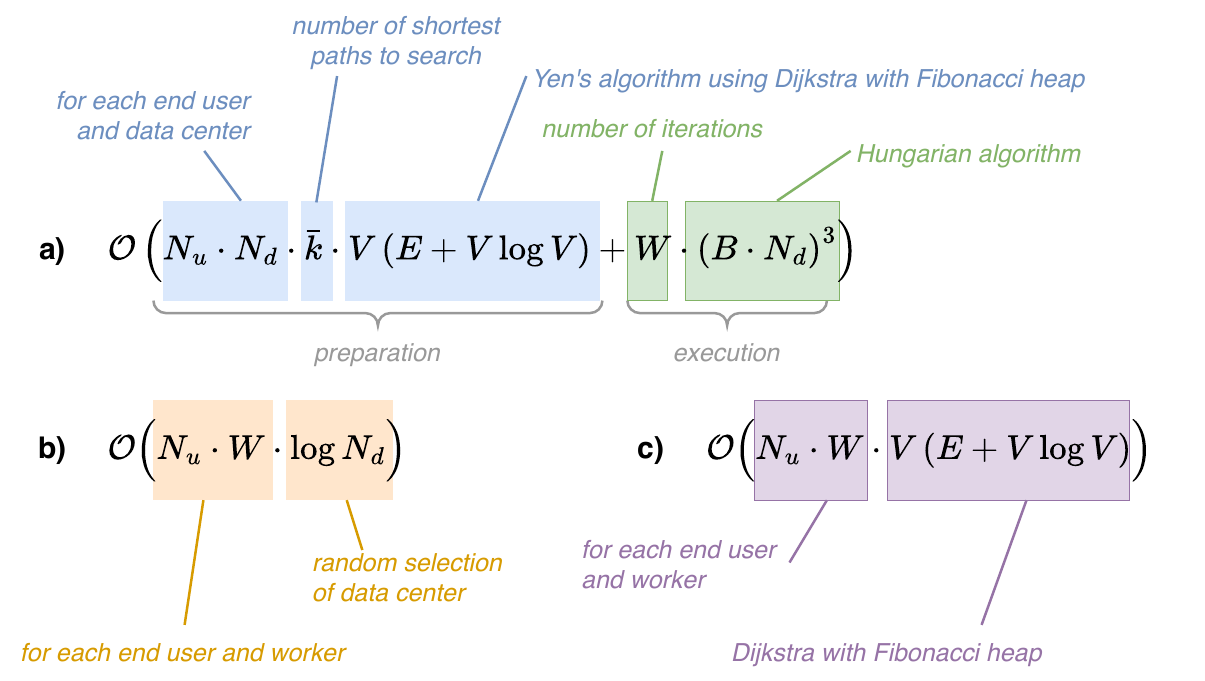}{Worst case time complexity of a) the load balancing algorithm to solve the \ac{QWAP} in \rsec{eval-fair:assignment} vs.\ the comparison algorithms b) random and c) shortest path.}

In this section we evaluate the load balancing algorithm defined in \rsec{eval-fair:assignment} using the simulator described in \rsec{summary:simulation}.
End users and data centers are selected randomly from the set of nodes $V$, with the following composition of the nodes: 10\% end users, 10\% data centers, 80\% intermediate nodes.
As comparison algorithms, we define:

\begin{myitemlist}
    \item \textit{Random}, which assigns each end user $W$ data centers at random, thus it maximizes fairness.
    \item \textit{Shortest path}, which assigns each end user the $W$ data centers that are closest in $G(E,V)$, in number of hops, thus it maximizes the net rate.
\end{myitemlist}

\noindent After the assignment, the resources are allocated using QDRR.

We report in \rfig{system-complexity} the worst case time complexity of the three algorithms, where we assume that the shortest path is computed using Dijkstra's algorithm with the help of a Fibonacci heap to keep edges sorted~\cite{fredman_fibonacci_1984}.
As can be seen, the load balancing algorithm is far more complex than random and shortest-path, in both the preparation and the execution phases; random, in particular, does not even depend on the graph size.
However, in the following we will see that the added complexity brings benefits that can be of potential interest to the future quantum network operators.

\myfigdoubleeps{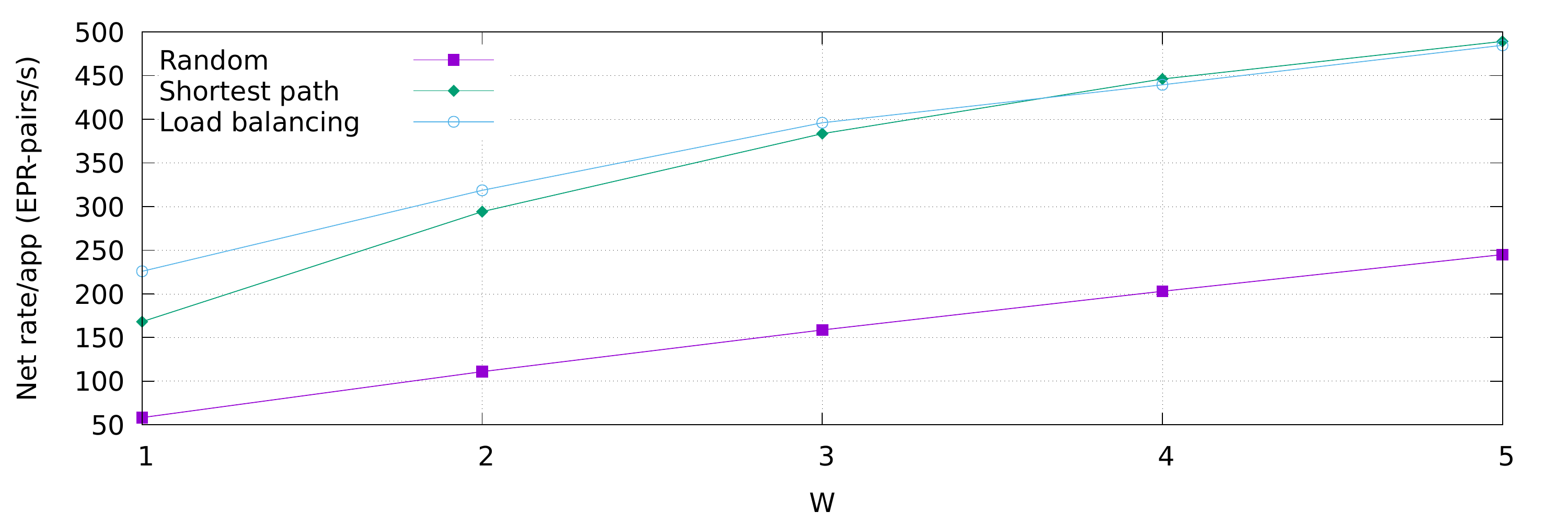}{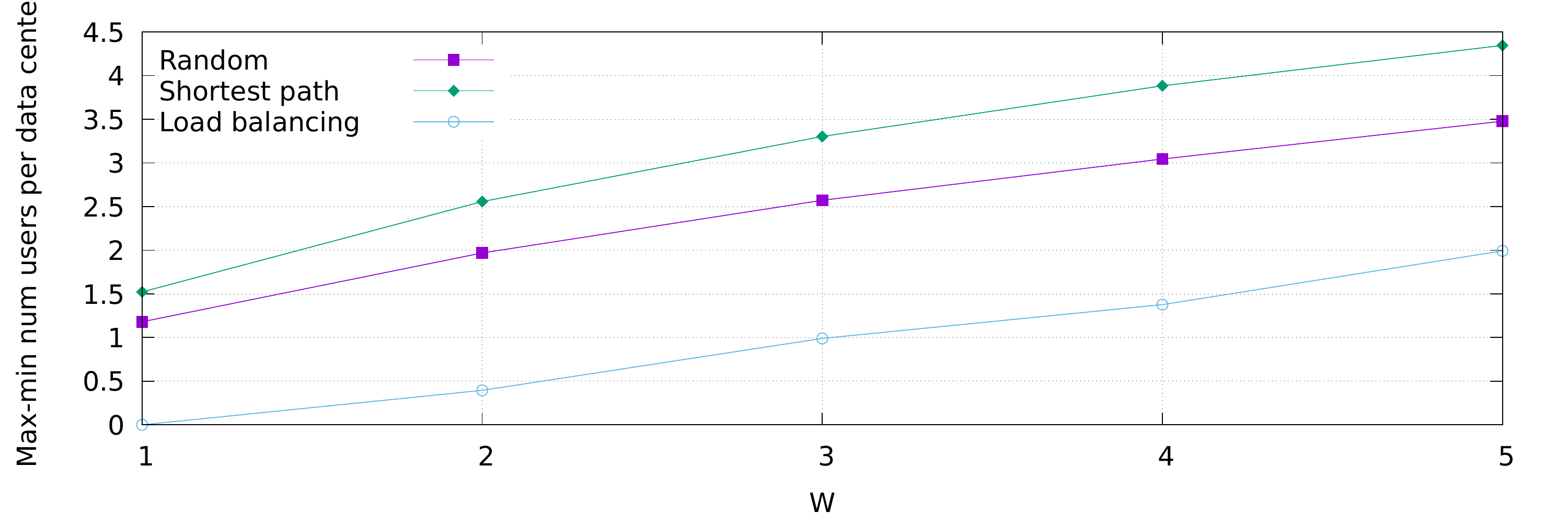}{005-vp-dense-10}{Net rate/app (top) and max-min number of users per data center (bottom) with random, shortest path, and load balancing, in a dense topology, with 10 apps, when increasing $W$ from 1 to 5.}

In \rfig{005-vp-dense-10} (top) we show the net rate/app with $N_u = 10$ and $W$ increasing from 1 to 5, in a dense topology.
As can be seen, the random algorithm performs poorly, because it does not consider at all the network topology.
On the other hand, shortest path and load balancing perform similarly, with the latter exhibiting a higher net rate with small values of $W$.
In \rfig{005-vp-dense-10} (bottom) we plot a measure of the spread of resources, as the max-min number of users per data center.
Load balancing performs consistently and significantly better than both random and shortest path, with the latter exhibiting the highest unfairness.
We note that our conclusions are limited to the settings of the scenarios simulated; in particular, different topologies or link capacity distributions can lead to situations where the gap between load balancing and either random or shortest path is reduced significantly.
However, a crucial advantage of our proposed solution, compared to its alternatives under test, is that it can adapt to different settings, thus it can perform well even when the scenario is not known or changes dynamically.

\myfigeps{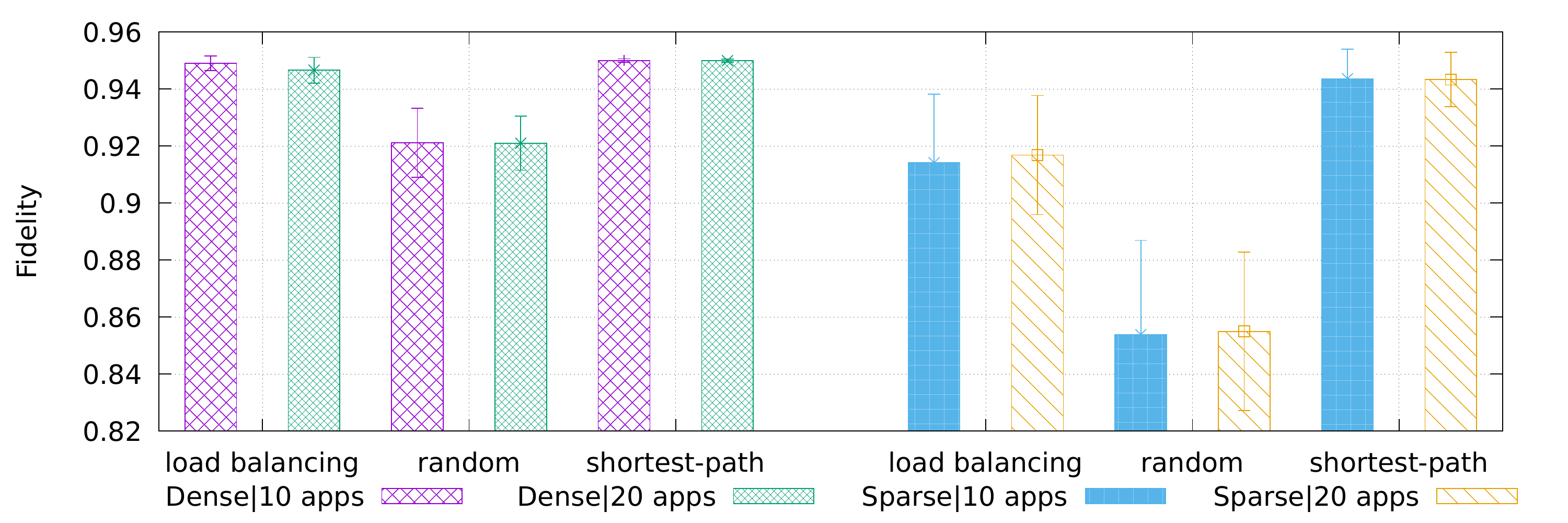}{Fidelity with random, shortest path, and load balancing, in dense vs.\ sparse topologies, with 10 vs.\ 20 apps and $W = 1$.}

The fidelity is shown in \rfig{005-vp-fidelity}, with $N_u = \{10,20\}$ and in dense vs.\ sparse topologies.
The number of end users/apps, i.e., $N_u$, does not affect the performance in a noticeable manner.
On the other hand, the fidelity is generally lower in sparse topologies, as expected.
Random performs worse, while load balancing and shortest path give similar results, with the former performing slightly worse only in the sparse case.
This can be explained by following the same line of reasoning for the net rate/app above. 


\myfigeps{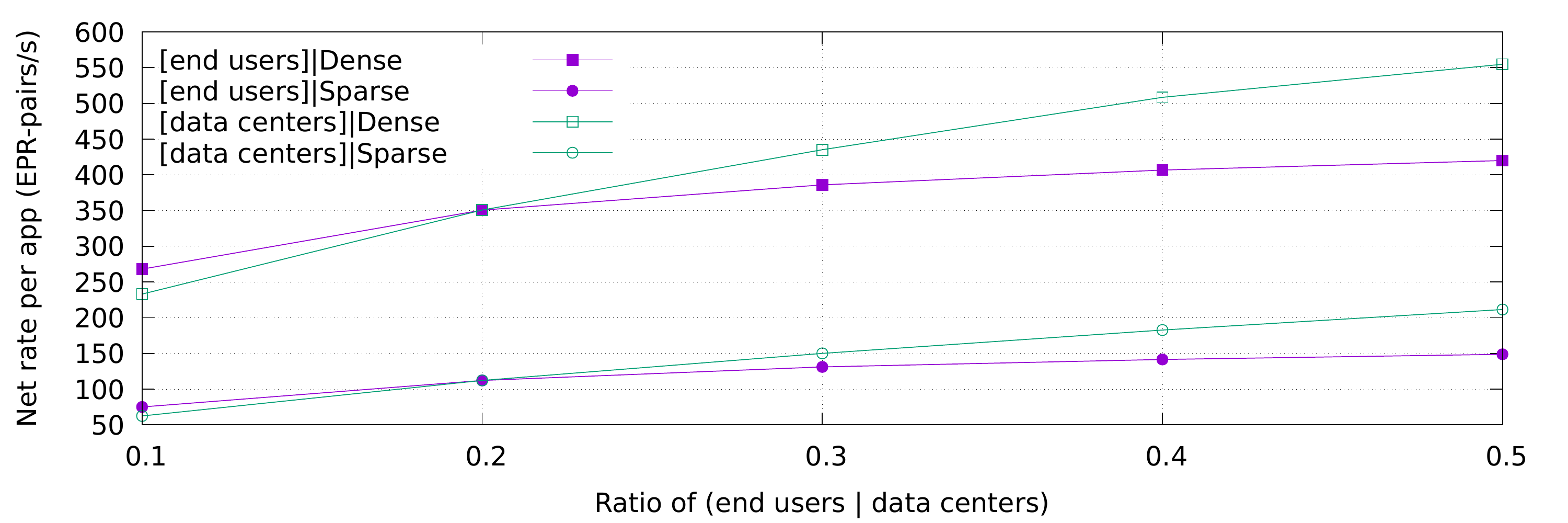}{Net rate/app with load balancing, in dense vs.\ sparse topologies, with 15 apps and $W = 3$, when increasing the fraction of nodes as either end users or data centers.}

To conclude the analysis, we modify the mix of nodes.
In one batch of simulations we increase the ratio of end users from 10\% (like in the results so far) to 50\%; in another one we do the same for data centers.
Results are shown only for load balancing in \rfig{005-mix-netrate-load-balancing}, in terms of the net rate, with $N_u = 15$ and $W = 3$.
As the ratio of data centers increases (green curves), the net rate/app increases almost linearly, as well.
On the other hand, increasing the number of users (blue curves), only provides a sub-linear performance improvement.
This is because the former case corresponds to increasing the physical resources provided to the users, while the latter only to a more uniform distribution of the same resources.
The conclusions are the same for dense and spare topologies, though the net rate for the latter is significantly lower.


\textit{In conclusion, assigning data centers to end users through the load balancing algorithm, which provides an approximate solution of the \ac{QWAP}, achieves an even distribution of resources, better than both random and shortest path assignment, without compromising on the net rate and fidelity of the end-to-end entanglement paths.}%

  \section{Conclusions}%
  \label{sec:conclusions}%
  In this paper we have studied two open issues in quantum networking for distributed quantum computing.
First, we have assessed through simulation the effectiveness of the QDRR resource allocation algorithm~\cite{cicconetti_resource_2022} in handling scenarios where applications have different priority weights or minimum fidelity requirements.
Second, we have defined a novel problem involving the selection of workers for a set of nodes hosting computation, called the \acf{QWAP}, which we have modeled as an optimization problem and for which we have proposed a heuristic called ``load balancing''.
The latter has been evaluated in comparison to alternatives seeking to maximize only fairness and the net rate of end-to-end entanglement, respectively, and the results have shown that load balancing achieves an excellent compromise in terms of the two metrics.
The source code and simulation scripts are publicly available to the community.

Further open research areas are: the use of purification
to increase fidelity at the expense of capacity; modeling distributed
\ac{QC} applications to understand their characteristic time scales
and requirements; integration with link layer protocols; incorporation in the simulation of more realistic models for the quantum channel and repeaters.%

\section*{Acknowledgment}

Work co-funded by EU, \textit{PON Ricerca e Innovazione} 2014--2020
FESR/FSC Project ARS01\_00734 QUANCOM, and European High-Performance
Computing Joint Undertaking (JU) under grant agreement No 101018180
HPCQS.
The paper reflects only the authors' view and the funding agencies
are not responsible for any use that may be made of its content.


\begin{thebibliography}{10}
\expandafter\ifx\csname url\endcsname\relax
  \def\url#1{\texttt{#1}}\fi
\expandafter\ifx\csname href\endcsname\relax
  \def\href#1#2{#2} \def\path#1{#1}\fi

\bibitem{preskill_quantum_2021}
J.~Preskill, \href{http://arxiv.org/abs/2106.10522}{Quantum computing 40 years
  later}, arXiv:2106.10522 [quant-ph]ArXiv: 2106.10522.

\bibitem{sevilla_forecasting_2020}
J.~Sevilla, C.~J. Riedel, \href{http://arxiv.org/abs/2009.05045}{Forecasting
  timelines of quantum computing}, arXiv:2009.05045 [quant-ph]ArXiv:
  2009.05045.

\bibitem{quantum_technology_and_application_consortium__qutac_industry_2021}
{Quantum Technology and Application Consortium – QUTAC}, A.~Bayerstadler, \textit{et al.}
  \href{https://epjquantumtechnology.springeropen.com/articles/10.1140/epjqt/s40507-021-00114-x}{Industry
  quantum computing applications}, EPJ Quantum Technology 8~(1) (2021) 25.
\newblock \href {http://dx.doi.org/10.1140/epjqt/s40507-021-00114-x}
  {\path{doi:10.1140/epjqt/s40507-021-00114-x}}.

\bibitem{gyongyosi_advances_2022}
L.~Gyongyosi, S.~Imre, \href{https://doi.org/10.1145/3524455}{Advances in the
  quantum internet}, Communications of the ACM 65~(8) (2022) 52--63.
\newblock \href {http://dx.doi.org/10.1145/3524455}
  {\path{doi:10.1145/3524455}}.

\bibitem{parekh_quantum_2021}
R.~Parekh, A.~Ricciardi, A.~Darwish, S.~DiAdamo,
  \href{http://arxiv.org/abs/2106.06841}{Quantum {Algorithms} and {Simulation}
  for {Parallel} and {Distributed} {Quantum} {Computing}}, arXiv:2106.06841
  [quant-ph]ArXiv: 2106.06841.

\bibitem{cicconetti_resource_2022}
C.~Cicconetti, M.~Conti, A.~Passarella, Resource {Allocation} in {Quantum}
  {Networks} for {Distributed} {Quantum} {Computing}, Proc. {IEEE} {SMARTCOMP} 2022.

\bibitem{shreedhar_efficient_1995}
M.~Shreedhar, G.~Varghese,
  \href{https://dl.acm.org/doi/10.1145/217391.217453}{Efficient fair queueing
  using {Deficit} {Round} {Robin}}, ACM SIGCOMM Computer Comm. Review 25~(4)
  (1995) 231--242.

\bibitem{muralidharan_optimal_2016}
S.~Muralidharan, L.~Li, J.~Kim, N.~Lütkenhaus, M.~D. Lukin, L.~Jiang,
  \href{http://www.nature.com/articles/srep20463}{Optimal architectures for
  long distance quantum communication}, Scientific Reports 6~(1) (2016) 20463.
\newblock \href {http://dx.doi.org/10.1038/srep20463}
  {\path{doi:10.1038/srep20463}}.

\bibitem{wang_field-deployable_2022}
Y.~Wang, A.~N. Craddock, R.~Sekelsky, M.~Flament, M.~Namazi,
  \href{http://arxiv.org/abs/2205.13091}{Field-deployable {Quantum} {Memory}
  for {Quantum} {Networking}}, Phys. Rev. Applied 18, 044058, 2022.

\bibitem{chakraborty_entanglement_2020}
K.~Chakraborty, D.~Elkouss, B.~Rijsman, S.~Wehner,
  \href{https://ieeexplore.ieee.org/document/9210823/}{Entanglement
  {Distribution} in a {Quantum} {Network}: {A} {Multicommodity} {Flow}-{Based}
  {Approach}}, IEEE Transactions on Quantum Engineering 1 (2020) 1--21.

\bibitem{sangouard_quantum_2011}
N.~Sangouard, C.~Simon, H.~de~Riedmatten, N.~Gisin,
  \href{https://link.aps.org/doi/10.1103/RevModPhys.83.33}{Quantum repeaters
  based on atomic ensembles and linear optics}, Reviews of Modern Physics
  83~(1) (2011) 33--80.
\newblock \href {http://dx.doi.org/10.1103/RevModPhys.83.33}
  {\path{doi:10.1103/RevModPhys.83.33}}.

\bibitem{briegel_quantum_1998}
H.-J. Briegel, W.~Dür, J.~I. Cirac, P.~Zoller,
  Quantum repeaters for communication, arXiv:quant-ph/9803056, 1998.

\bibitem{dai_optimal_2020}
W.~Dai, T.~Peng, M.~Z. Win,
  \href{https://ieeexplore.ieee.org/document/8967073/}{Optimal {Remote}
  {Entanglement} {Distribution}}, IEEE Journal on Selected Areas in
  Communications 38~(3) (2020) 540--556.
\newblock \href {http://dx.doi.org/10.1109/JSAC.2020.2969005}
  {\path{doi:10.1109/JSAC.2020.2969005}}.

\bibitem{bouwmeester_experimental_1997}
D.~Bouwmeester, J.-W. Pan, K.~Mattle, M.~Eibl, H.~Weinfurter, A.~Zeilinger,
  \href{http://www.nature.com/articles/37539}{Experimental quantum
  teleportation}, Nature 390~(6660) (1997) 575--579.
\newblock \href {http://dx.doi.org/10.1038/37539} {\path{doi:10.1038/37539}}.

\bibitem{van_meter_path_2013}
R.~Van~Meter, T.~Satoh, T.~D. Ladd, W.~J. Munro, K.~Nemoto,
  \href{http://arxiv.org/abs/1206.5655}{Path {Selection} for {Quantum}
  {Repeater} {Networks}}, Networking Science 3~(1-4) (2013) 82--95, arXiv:
  1206.5655.

\bibitem{caleffi_optimal_2017}
M.~Caleffi, \href{http://ieeexplore.ieee.org/document/8068178/}{Optimal
  {Routing} for {Quantum} {Networks}}, IEEE Access 5 (2017) 22299--22312.
\newblock \href {http://dx.doi.org/10.1109/ACCESS.2017.2763325}
  {\path{doi:10.1109/ACCESS.2017.2763325}}.

\bibitem{chakraborty_distributed_2019}
K.~Chakraborty, F.~Rozpedek, A.~Dahlberg, S.~Wehner,
  \href{http://arxiv.org/abs/1907.11630}{Distributed {Routing} in a {Quantum}
  {Internet}}, [quant-ph]ArXiv: 1907.11630.

\bibitem{pant_routing_2019}
M.~Pant, H.~Krovi, D.~Towsley, L.~Tassiulas, L.~Jiang, P.~Basu, D.~Englund,
  S.~Guha, \href{http://www.nature.com/articles/s41534-019-0139-x}{Routing
  entanglement in the quantum internet}, npj Quantum Information 5~(1) (2019)
  25.
\newblock \href {http://dx.doi.org/10.1038/s41534-019-0139-x}
  {\path{doi:10.1038/s41534-019-0139-x}}.

\bibitem{cicconetti_request_2021}
C.~Cicconetti, M.~Conti, A.~Passarella, Request {Scheduling} in {Quantum}
  {Networks}, IEEE Transactions on Quantum Engineering 2 (2021) 2--17,

\bibitem{li_effective_2021}
C.~Li, T.~Li, Y.-X. Liu, P.~Cappellaro,
  \href{http://www.nature.com/articles/s41534-020-00344-4}{Effective routing
  design for remote entanglement generation on quantum networks}, npj Quantum
  Information 7~(1) (2021) 10.
\newblock \href {http://dx.doi.org/10.1038/s41534-020-00344-4}
  {\path{doi:10.1038/s41534-020-00344-4}}.

\bibitem{van_meter_quantum_2021}
R.~Van~Meter, R.~Satoh, N.~Benchasattabuse, T.~Matsuo, M.~Hajdušek, T.~Satoh,
  S.~Nagayama, S.~Suzuki, A {Quantum}
  {Internet} {Architecture}, Proc. {IEEE} {QCE} 2022, pp. 341--352.

\bibitem{zhao_quantum_2021}
Y.~Zhao, C.~Qiao, \href{http://arxiv.org/abs/2105.08109}{Quantum {Transport}
  {Protocols} for {Distributed} {Quantum} {Computing}}, arXiv:2105.08109, 2021.

\bibitem{dahlberg_link_2019}
A.~Dahlberg, M.~Skrzypczyk, T.~Coopmans, L.~Wubben, F.~Rozpędek, M.~Pompili,
  A.~Stolk, P.~Pawełczak, R.~Knegjens, J.~de~Oliveira~Filho, R.~Hanson,
  S.~Wehner, \href{https://dl.acm.org/doi/10.1145/3341302.3342070}{A link layer
  protocol for quantum networks}, Proc. {ACM} {SIGCOMM} 2019, pp.
  159--173.

\bibitem{cuomo_optimized_2021}
D.~Cuomo, M.~Caleffi, K.~Krsulich, F.~Tramonto, G.~Agliardi, E.~Prati, A.~S.
  Cacciapuoti, Optimized compiler for {Distributed} {Quantum} {Computing}, {ACM} Trans. on Quantum Computing, 2023 (to appear).

\bibitem{dahlberg_netqasmlow-level_2022}
A.~Dahlberg, B.~v.~d. Vecht, C.~D. Donne, M.~Skrzypczyk, I.~t. Raa,
  W.~Kozlowski, S.~Wehner,
  \href{https://doi.org/10.1088/2058-9565/ac753f}{{NetQASM}—a low-level
  instruction set architecture for hybrid quantum–classical programs in a
  quantum internet}, Quantum Science and Technology 7~(3) (2022).

\bibitem{van_meter_system_2009}
R.~Van~Meter, T.~Ladd, W.~Munro, K.~Nemoto,
  \href{http://ieeexplore.ieee.org/document/4695947/}{System {Design} for a
  {Long}-{Line} {Quantum} {Repeater}}, IEEE/ACM Trans. on Networking
  17~(3) (2009).

\bibitem{zhao_e2e_2022}
Y.~Zhao, G.~Zhao, C.~Qiao, {E2E} {Fidelity} {Aware} {Routing} and
  {Purification} for {Throughput} {Maximization} in {Quantum} {Networks}, Proc.
  {IEEE} {INFOCOM} 2022, pp. 480--489.

\bibitem{pompili_realization_2021}
M.~Pompili, S.~L.~N. Hermans, S.~Baier, H.~K.~C. Beukers, P.~C. Humphreys,
  R.~N. Schouten, R.~F.~L. Vermeulen, M.~J. Tiggelman, L.~d.~S. Martins,
  B.~Dirkse, S.~Wehner, R.~Hanson,
  Realization of a multi-node quantum network of remote solid-state qubits, Science 372~(6539) (2021) 259--264.

\bibitem{patil_entanglement_2022}
A.~Patil, M.~Pant, D.~Englund, D.~Towsley, S.~Guha,
  \href{https://www.nature.com/articles/s41534-022-00536-0}{Entanglement
  generation in a quantum network at distance-independent rate}, npj Quantum
  Information 8~(1) (2022).

\bibitem{jalali_data_2000}
A.~Jalali, R.~Padovani, R.~Pankaj, Data throughput of {CDMA}-{HDR} a high
  efficiency-high data rate personal communication wireless system, Proc. {IEEE}
  {VTC2000}-{Spring} 2000, pp. 1854--1858 vol.3.

\bibitem{martello_bound_1981}
S.~Martello, P.~Toth,
  \href{https://www.sciencedirect.com/science/article/pii/0166218X81900056}{A
  {Bound} and {Bound} algorithm for the zero-one multiple knapsack problem},
  Discrete Applied Mathematics 3~(4) (1981) 275--288.

\bibitem{kuhn_hungarian_1955}
H.~W. Kuhn,
  \href{https://onlinelibrary.wiley.com/doi/abs/10.1002/nav.3800020109}{The
  {Hungarian} method for the assignment problem}, Naval Research Logistics
  Quarterly 2~(1-2) (1955) 83--97.

\bibitem{yen_finding_1971}
J.~Y. Yen,
  \href{https://pubsonline.informs.org/doi/abs/10.1287/mnsc.17.11.712}{Finding
  the {K} {Shortest} {Loopless} {Paths} in a {Network}}, Management Science
  17~(11) (1971) 712--716.

\bibitem{fredman_fibonacci_1984}
M.~Fredman, R.~Tarjan, Fibonacci {Heaps} {And} {Their} {Uses} {In} {Improved}
  {Network} {Optimization} {Algorithms}, in: 25th {Annual} {Symposium}
  {onFoundations} of {Computer} {Science}, J. ACM 34, 3 (July 1987), 596–615.

\end{thebibliography}



\begin{acronym}
  \acro{3GPP}{Third Generation Partnership Project}
  \acro{5G-PPP}{5G Public Private Partnership}
  \acro{AA}{Authentication and Authorization}
  \acro{ADF}{Azure Durable Function}
  \acro{AI}{Artificial Intelligence}
  \acro{API}{Application Programming Interface}
  \acro{AP}{Access Point}
  \acro{AR}{Augmented Reality}
  \acro{BGP}{Border Gateway Protocol}
  \acro{BSP}{Bulk Synchronous Parallel}
  \acro{BS}{Base Station}
  \acro{CDF}{Cumulative Distribution Function}
  \acro{CFS}{Customer Facing Service}
  \acro{CPU}{Central Processing Unit}
  \acro{DAG}{Directed Acyclic Graph}
  \acro{DHT}{Distributed Hash Table}
  \acro{DNS}{Domain Name System}
  \acro{DRR}{Deficit Round Robin} 
  \acro{ETSI}{European Telecommunications Standards Institute}
  \acro{FCFS}{First Come First Serve}
  \acro{FSM}{Finite State Machine}
  \acro{FaaS}{Function as a Service}
  \acro{GPU}{Graphics Processing Unit}
  \acro{HTML}{HyperText Markup Language}
  \acro{HTTP}{Hyper-Text Transfer Protocol}
  \acro{ICN}{Information-Centric Networking}
  \acro{IETF}{Internet Engineering Task Force}
  \acro{IIoT}{Industrial Internet of Things}
  \acro{ILP}{Integer Linear Programming}
  \acro{IPP}{Interrupted Poisson Process}
  \acro{IP}{Internet Protocol}
  \acro{ISG}{Industry Specification Group}
  \acro{ITS}{Intelligent Transportation System}
  \acro{ITU}{International Telecommunication Union}
  \acro{IT}{Information Technology}
  \acro{IaaS}{Infrastructure as a Service}
  \acro{IoT}{Internet of Things}
  \acro{JSON}{JavaScript Object Notation}
  \acro{K8s}{Kubernetes}
  \acro{KVS}{Key-Value Store}
  \acro{LCM}{Life Cycle Management}
  \acro{LL}{Link Layer}
  \acro{LOCC}{Local Operations and Classical Communication}
  \acro{LTE}{Long Term Evolution}
  \acro{MAC}{Medium Access Layer}
  \acro{MBWA}{Mobile Broadband Wireless Access}
  \acro{MCC}{Mobile Cloud Computing}
  \acro{MEC}{Multi-access Edge Computing}
  \acro{MEH}{Mobile Edge Host}
  \acro{MEPM}{Mobile Edge Platform Manager}
  \acro{MEP}{Mobile Edge Platform}
  \acro{ME}{Mobile Edge}
  \acro{ML}{Machine Learning}
  \acro{MNO}{Mobile Network Operator}
  \acro{NAT}{Network Address Translation}
  \acro{NISQ}{Noisy Intermediate-Scale Quantum}
  \acro{NFV}{Network Function Virtualization}
  \acro{NFaaS}{Named Function as a Service}
  \acro{OSPF}{Open Shortest Path First}
  \acro{OSS}{Operations Support System}
  \acro{OS}{Operating System}
  \acro{OWC}{OpenWhisk Controller}
  \acro{PMF}{Probability Mass Function}
  \acro{PPP}{Poisson Point Process}
  \acro{PU}{Processing Unit}
  \acro{PaaS}{Platform as a Service}
  \acro{PoA}{Point of Attachment}
  \acro{PPP}{Poisson Point Process}
  \acro{QC}{Quantum Computing}
  \acro{QKD}{Quantum Key Distribution}
  \acro{QoE}{Quality of Experience}
  \acro{QoS}{Quality of Service}
  \acro{QWAP}{Quantum Workers' Assignment Problem}
  \acro{RPC}{Remote Procedure Call}
  \acro{RR}{Round Robin}
  \acro{RSU}{Road Side Unit}
  \acro{SBC}{Single-Board Computer}
  \acro{SDK}{Software Development Kit}
  \acro{SDN}{Software Defined Networking}
  \acro{SJF}{Shortest Job First}
  \acro{SLA}{Service Level Agreement}
  \acro{SMP}{Symmetric Multiprocessing}
  \acro{SoC}{System on Chip}
  \acro{SLA}{Service Level Agreement}
  \acro{SRPT}{Shortest Remaining Processing Time}
  \acro{SPT}{Shortest Processing Time}
  \acro{STL}{Standard Template Library}
  \acro{SaaS}{Software as a Service}
  \acro{TCP}{Transmission Control Protocol}
  \acro{TSN}{Time-Sensitive Networking}
  \acro{UDP}{User Datagram Protocol}
  \acro{UE}{User Equipment}
  \acro{URI}{Uniform Resource Identifier}
  \acro{URL}{Uniform Resource Locator}
  \acro{UT}{User Terminal}
  \acro{VANET}{Vehicular Ad-hoc Network}
  \acro{VIM}{Virtual Infrastructure Manager}
  \acro{VR}{Virtual Reality}
  \acro{VM}{Virtual Machine}
  \acro{VNF}{Virtual Network Function}
  \acro{WLAN}{Wireless Local Area Network}
  \acro{WMN}{Wireless Mesh Network}
  \acro{WRR}{Weighted Round Robin}
  \acro{YAML}{YAML Ain't Markup Language}
\end{acronym}

\end{document}